\documentclass{PoS}

\def\tgs{$\gamma_S$}

\def\sqrtsnnt{$\sqrt{s}_\mathrm{NN}$}

\def\EperNt{$\langle E \rangle / \langle N \rangle$ }
\def\tpm{$\pm$}

\def\detahm{$\mathrm{d}N_{h^-}/\mathrm{d}\eta$}
\def\agev{$A$ GeV}
\def\agevt{$A$ GeV }

\newcommand{\beq}{\begin{equation}}
\newcommand{\eeq}{\end{equation}}
\newcommand{\bq}{\begin{eqnarray}}
\newcommand{\eq}{\end{eqnarray}}

\title{Statistical hadronization model description for 
rapidity densities at RHIC}

\ShortTitle{Statistical hadronization model description for...}

\author{\speaker{Jaakko Manninen}\\ %\thanks{A footnote may follow.}\\
        Department of Physical Sciences, University of Oulu, Oulu, Finland\\
        E-mail: \email{jaakko.manninen@oulu.fi}}

\abstract{The rapidity densities at \sqrtsnnt~= 200 and 130 GeV measured 
at Relativistic Heavy-Ion Collider by STAR and PHENIX collaborations are 
analysed within the statistical hadronization model at chemical 
freeze-out. We find that the model can describe the experimental 
rapidity densities well. The corresponding chemical freeze-out parameters
are determined and they are seen to be in agreement with what we expect from 
our previous analyses at lower beam energies at AGS and SPS.}
\FullConference{The 3rd edition of the International Workshop --- The Critical Point and Onset of Deconfinement --- \\
		 July 3-7 2006\\
		 Galileo Galilei Institute, Florence, Italy}

\begin{document}

\section{Introduction}

We analyze within the statistical hadronization model
the rapidity densities at mid-rapidity in the central Au-Au
collisions at \sqrtsnnt=200 GeV and \sqrtsnnt=130 GeV measured by 
STAR and PHENIX 
collaborations~\cite{Adams:2003xp,Adams:2004ux,
Adler:2001yq,Adler:2002uv,Adler:2002wn,Adler:2002xv,Adams:2003ve,Adams:2003yh,
Adams:2003fy,Adams:2006ke,Adcox:2002au,Adcox:2003nr}. 
The analysis is carried out in the grand canonical ensemble
with the supplementary \tgs~parameter. This version of the statistical 
hadronization model is described in detail 
elsewhere~\cite{Becattini:2003wp,Becattini:2005xt}.

Statistical model analyses at mid-rapidity at RHIC are often performed 
by fitting the chemical freeze-out parameters to the ratios of measured
rapidity 
densities~\cite{Braun-Munzinger:2001ip,Baran:2003nm,Cleymans:2004pp,Dumitru:2005hr,Andronic:2005yp}. 
It is easy to see that in a boost invariant scenario
the ratios of rapidity densities represent the ratios of the multiplicities 
since
\beq
\frac{N_A}{N_B} = \frac{\int (dN_A/dy) dy}{\int (dN_B/dy) dy}
= \frac{(dN_A/dy) \int dy}{(dN_B/dy) \int dy} 
= \frac{dN_A/dy}{dN_B/dy} 
\eeq
in which the second step results from the fact that the rapidity densities,
being boost invariant, are independent on rapidity. We adopt a 
different approach here and do not include any ratios of rapidity densities
in the analysis as this can lead to a large bias in the resulting best fit 
parameters~\cite{Francesco}. 
Instead, 
as in our statistical model analyses at 
AGS and SPS~\cite{Becattini:2003wp,Becattini:2005xt}, we fit  
a scaling parameter $V'$, common for all particle species, characterizing
the size of the hadron emitting source in the limited rapidity window around 
$y$=0. 

We have shown before~\cite{Becattini:2003wp} that performing 
statistical analysis in a limited rapidity window at SPS energies can 
artificially enhance the relative strangeness content in the system. 
This is due to the fact that heavier particles have more narrow rapidity 
distributions and thus a kinematic cut around mid-rapidity may lead to a 
situation in which relatively larger fraction of massive particles is 
taken into account. However, if a sufficiently wide Bjorken type of 
boost invariant region, much wider than the typical width of 
the pion rapidity distribution coming from a single cluster at kinetic 
freeze-out is formed around mid-rapidity, then the rapidity densities 
of different particle species are approximately constant in this region. 
In this case it might be possible to determine the characteristics of an 
average fireball at mid-rapidity without introducing a bias arising from the 
kinematic cut~\cite{Becattini:2005xt}. 

\section{Analysis and results}

The rapidity densities in 0-5\% most central Au-Au collisions 
measured by STAR 
at \sqrtsnnt=200 GeV and statistical hadronization model 
predictions for the same rapidity densities along our best fit 
parameters are shown in Table~\ref{star200}. 
All other rapidity densities are corrected 
for the weak decay contamination except proton and antiproton rapidity
densities which include the weak decay products of $\Lambda$ and 
$\bar{\Lambda}$.
Fit quality is generally good and rapidity densities are well described 
within the statistical hadronization model. Our findings agree with
the similar analysis performed recently by the STAR 
collaboration~\cite{Adams:2006ke}.

%%%%%%%%%%%%%%%%%%%%%%%%%%%%%%%%%%%%%%%%%%%%%%%%%%%%%%%%%%%%%%%%%
%%% STAR 200
%%%%%%%%%%%%%%%%%%%%%%%%%%%%%%%%%%%%%%%%%%%%%%%%%%%%%%%%%%%%%%%%%
%
%
%
\begin{table}
\begin{center}
\begin{tabular}{|c|c|c|}
\hline
  particle  &     dN/dy        & $\;\;\;\;\;\;$SHM$\;\;\;\;\;\;$  \\
\hline
$\pi^+$~\cite{Adams:2003xp} &  322\tpm32        & 325.0 \\
$\pi^-$~\cite{Adams:2003xp} &  327\tpm33        & 327.4 \\
$K^+$~\cite{Adams:2003xp}   &  51.3\tpm7.7      & 57.1  \\
$K^-$~\cite{Adams:2003xp}   &  49.5\tpm7.4      & 53.5  \\
$p$~\cite{Adams:2003xp}
%\footnote{Includes weak feeddown contibution from $\Lambda$.} 
                                 &  34.7\tpm6.2      & 42.9  \\
$\bar{p}$~\cite{Adams:2003xp}
%\footnote{Includes weak feeddown contibution from $\bar{\Lambda}$.} 
                                 &  26.7\tpm4.0      & 30.9  \\
$\phi$~\cite{Adams:2004ux}       &  7.70\tpm0.30\tpm0.85     & 7.10 \\
$\Lambda$~\cite{Adams:2006ke}    &  16.7\tpm0.2\tpm1.1       & 16.0 \\
$\bar{\Lambda}$~\cite{Adams:2006ke}
                                 &  12.7\tpm0.2\tpm0.9       & 12.1 \\
$\Xi^-$~\cite{Adams:2006ke}      &  2.17\tpm0.06\tpm0.19     & 1.87 \\
$\bar{\Xi}^+$~\cite{Adams:2006ke}&  1.83\tpm0.05\tpm0.20     & 1.53 \\
$\Omega + \bar{\Omega}$~\cite{Adams:2006ke}
                                 &  0.53\tpm0.04\tpm0.04     & 0.63 \\
\hline
\hline
$V'T^3e^{-0.7/T}$  &    \multicolumn{2}{|c|}{12.5\tpm0.7}  \\
$T$ [MeV]          &    \multicolumn{2}{|c|}{161.0\tpm3.9}  \\
$\mu_B$ [MeV]      &    \multicolumn{2}{|c|}{30.0\tpm9.8}  \\
$\gamma_S$         &    \multicolumn{2}{|c|}{1.02\tpm0.05}  \\
$\chi^2/dof$       &    \multicolumn{2}{|c|}{12.5/8} \\
\EperNt [GeV]      &    \multicolumn{2}{|c|}{0.99\tpm0.04} \\
\hline
\end{tabular}
\caption{{\bf Top panel:} STAR rapidity densities in Au-Au collisions at 
\sqrtsnnt=200 GeV in
the 0-5\% most central collisions. Contribution 
from weak decays have been subtracted for all particle species except that 
proton and antiproton rapidity densities include weak decay products of 
$\Lambda$ and $\bar{\Lambda}$.  Errors quoted 
are statistical + systematic and are taken into account in quadrature. 
For pions, kaons and nucleons, statistical errors are negligible and 
thus are not shown. 
{\bf Bottom panel:} The statistical hadronization model best fit parameters
at chemical freeze-out and the mean energy per particle, averaged over all 
particle species, determined from the above experimental data.
}\label{star200}
\end{center}
\end{table}

The STAR data in Au-Au collisions at \sqrtsnnt=130 GeV needs to be 
somewhat manipulated, since some
of the rapidity densities are measured in different centrality windows.
In order to keep consistency with the STAR 200$A$ GeV data and PHENIX 130$A$ 
GeV data, we have chosen to extrapolate all rapidity densities into the 
same centrality selection, namely to the 0-5\% most central collisions.
This is done by assuming that all final hadronic rapidity densities 
scale linearly 
\begin{equation}\label{linx}
\frac{dN}{dy} = a + b\frac{dN_{h^-}}{d\eta}
\end{equation}
with the negative hadron pseudorapidity density \detahm~in central and 
semi-central collisions. This is approximately true in the boost invariant
scenario around mid-rapidity provided that soft processes dominate the 
particle production and all particle multiplicities are proportional 
to the negative hadron multiplicity. This seems to hold true relatively 
well for all the measured particle species.

Using the STAR inelastic cross section versus \detahm~data 
from~\cite{Adler:2001yq}, we have calculated the 
negative hadron pseudorapidity densities
\detahm~in several centrality windows in the central pseudorapidity region 
($|\eta|<0.5$) in order to 
estimate the rapidity densities for each of the particle species in the 
most central collision window. Particularly, we have calculated
\detahm~= 297 in the 0-5\% most central collisions. This is in good 
agreement with STAR collaboration estimate 
\detahm~$\approx$ 300\tpm6\%~\cite{Adams:2003yh,Adler:2002uv} 
and we have chosen the same 6\% as our error leading to the 
\detahm~= 297\tpm18 in the 0-5\% most central collisions.

Pions and lambdas are already measured in the 0-5\% most central collisions 
and 
need no manipulation while the rapidity densities for other particle 
species need to be extrapolated. The negative hadron pseudorapidity 
densities as a function of centrality are published along 
kaon rapidity densities~\cite{Adler:2002wn} in various centrality windows
and thus we have fitted Eq. (\ref{linx}) to the data and used 
\detahm~= 297\tpm18 to estimate the $K^\pm$ rapidity densities in the 0-5\% 
most 
central collisions. Protons and antiprotons~\cite{Adams:2003ve} are
measured in the same centrality windows as kaons and thus we have used the 
same negative hadron
pseudorapidity densities as for kaons in order to estimate the $p$
and $\bar{p}$ rapidity densities in the 0-5\% most central collisions.

$\Xi^-$ and $\Xi^+$~\cite{Adams:2003fy} as well as $\phi$~\cite{Adler:2002xv}
are measured in three different 
centrality windows and similarly to kaons and nucleons, we have 
estimated rapidity densities in the 0-5\% most central collisions 
by assuming linear 
dependence on the negative hadron pseudorapidity density. Unlike in the case
of kaons and nucleons, the 
negative hadron pseudorapidity densities estimated by STAR collaboration 
in these windows deviate 
somewhat from the ones we have calculated. In order to keep the same 
overall scale, we have used our calculated pseudorapidity spectrum to 
estimate the 
$\Xi^\pm$ and $\phi$ rapidity densities in the 0-5\% most central collisions. 
Also in~\cite{Adams:2003fy}, the 0-10\% most central rapidity density 
for $\Omega + \bar{\Omega}$
is given and without better knowledge, we have estimated the 
$\Omega + \bar{\Omega}$ rapidity density in the 0-5\% most central 
collisions by scaling the yields with a factor 
$(dN_{h^-}/d\eta)_{5\%}/(dN_{h^-}/d\eta)_{10\%}$=1.06.
No weak decay corrections have been applied to any of the rapidity densities
except $\pi^\pm$ densities do not include weak decay products of 
$\Lambda$, $\bar{\Lambda}$ and $K^0_s$.
%%%%%%%%%%%%%%%%%%%%%%%%%%%%%%%%%%%%%%%%%%%%%%%%%%%%%%%%%%%%%%%%%
%%% STAR 130
%%%%%%%%%%%%%%%%%%%%%%%%%%%%%%%%%%%%%%%%%%%%%%%%%%%%%%%%%%%%%%%%%
%
\begin{table}[!t]
\begin{center}
\begin{tabular}{|c|c|c|}
\hline
particle   & dN/dy   & $\;\;\;\;\;\;$SHM$\;\;\;\;\;\;$ \\
\hline

$\pi^+$~\cite{Adams:2003yh}
             & 239.0\tpm3.0\tpm2.0\tpm10.0        & 233.8 \\
$\pi^-$~\cite{Adams:2003yh}
             & 239.0\tpm3.0\tpm2.0\tpm10.0        & 236.6 \\
$\Lambda$~\cite{Adler:2002uv}
             &  17.20\tpm0.4\tpm1.72              & 17.98 \\
$\bar{\Lambda}$~\cite{Adler:2002uv} 
             &  12.3\tpm0.3\tpm1.23               & 12.5  \\      
\hline
$K^+$%~\cite{Adler:2002wn}
             & 45.75\tpm0.6\tpm6.0\tpm2.85        & 46.92   \\      
$K^-$%~\cite{Adler:2002wn}
             & 43.16\tpm0.6\tpm5.4\tpm2.57        & 43.36   \\   
$p$%~\cite{Adams:2003ve}
             & 26.15\tpm0.23\tpm5.8\tpm1.63       & 32.17   \\      
$\bar{p}$%~\cite{Adams:2003ve}
             & 18.85\tpm0.16\tpm4.12\tpm1.15      & 21.96   \\      
$\Xi^-$%~\cite{Adams:2003fy} 
             & 2.13\tpm0.14\tpm0.20\tpm0.11       & 1.80    \\      
$\bar{\Xi}^+$%~\cite{Adams:2003fy}   
             & 1.78\tpm0.12\tpm0.17\tpm0.12       & 1.42    \\      
$\Omega+\bar{\Omega}$%~\cite{Adams:2003fy}
             &  0.586\tpm0.11\tpm0.056\tpm0.035   & 0.702   \\
$\phi$%~\cite{Adler:2002xv}
             & 6.09\tpm0.37\tpm0.69\tpm0.34       & 6.75    \\      
\hline
\hline
$V'T^3e^{-0.7/T}$     & \multicolumn{2}{|c|}{8.46\tpm0.45}   \\
$T$ [MeV]            & \multicolumn{2}{|c|}{160.5\tpm4.3}   \\
$\mu_B$ [MeV]        & \multicolumn{2}{|c|}{35.6\tpm12.8}   \\
$\gamma_S$           & \multicolumn{2}{|c|}{1.20\tpm0.08}   \\
$\chi^2/dof$         & \multicolumn{2}{|c|}{7.0/8}          \\
\EperNt [GeV]        & \multicolumn{2}{|c|}{1.00\tpm0.04}   \\
\hline
\end{tabular}
\caption{{\bf Top panel:} Estimated rapidity densities in Au-Au collisions 
at \sqrtsnnt=130 GeV in the 0-5\% most central collisions measured by STAR 
collaboration. Rapidity densities of pions and lambdas
are measurements while rapidity densities of other 
particle species are our estimates. 
Errors for our estimates are 
statistical + systematic + our extrapolation and are taken into account in
quadrature. Statistical and systematic 
errors quoted are the experimental errors in the most central experimentally 
accessible window, which are 0-6\% for $K^\pm$, $p$ and $\bar{p}$, 
0-10\% for $\Xi^\pm$ and $\Omega + \bar{\Omega}$ and 0-11\% for $\phi$.
No weak decay corrections have been applied except $\pi^\pm$ 
rapidity densities do not include weak decay products of 
$\Lambda$, $\bar{\Lambda}$ and $K^0_s$. 
{\bf Bottom panel:} The statistical hadronization model best fit parameters
at chemical freeze-out and the mean energy per particle
determined from the above experimental data.\label{star130}
}
\end{center}
\end{table}

Table \ref{star130} shows the extrapolated (experimental) and the 
statistical hadronization model predictions for the rapidity
densities in the 0-5\% most central collisions as well as the statistical model
best fit parameters in Au-Au collisions at \sqrtsnnt=130 GeV. Again, the 
fit quality
is good and most of the rapidity densities are described well within the 
statistical hadronization model. The chemical freeze-out
temperature is about the same and baryon chemical potential is few MeV 
higher than in central Au-Au collisions at 200\agev. Unexpectedly the 
strangeness under-saturation parameter $\gamma_S$ seems to be significantly
over unity. 
To cross check our results, we can analyze the rapidity densities 
in central Au-Au collisions
at the same center-of-mass energy \sqrtsnnt=130 GeV 
measured by the PHENIX collaboration. PHENIX has measured a subset of the
particle species measured by STAR, namely $\pi^\pm$, 
$K^\pm$, $p$, $\bar{p}$, $\Lambda$ and 
$\bar{\Lambda}$~\cite{Adcox:2003nr,Adcox:2002au}. 
Our extrapolated rapidity densities agree with the corresponding
PHENIX values
(see Tables~\ref{star130} and~\ref{phenix130}). No weak decay corrections have been 
applied to the PHENIX data which is the reason why $\pi^\pm$ rapidity 
densities deviate among PHENIX and STAR data at the same beam energy.
In all cases, strong and weak decays corresponding 
to the experimental definitions are taken carefully into account in our
analyses.
The statistical model best fit parameters estimated from the PHENIX data 
agree with the ones coming from a fit to the larger data 
sample measured by STAR. This rules out the possibility that our 
extrapolations with STAR data would bias the fit towards 
unexpectedly large $\gamma_S$.

%%%%%%%%%%%%%%%%%%%%%%%%%%%%%%%%%%%%%%%%%%%%%%%%%%%%%%%%%%%%%%%%%
%%% PHENIX 130
%%%%%%%%%%%%%%%%%%%%%%%%%%%%%%%%%%%%%%%%%%%%%%%%%%%%%%%%%%%%%%%%%
\begin{table}[!t]
\begin{center}
\begin{tabular}{|c|c|c|}
\hline
  particle    & dN/dy        & $\;\;\;\;\;\;$SHM$\;\;\;\;\;\;$  \\
\hline
$\pi^+$         & 276\tpm3\tpm36               &   264.4 \\
$\pi^-$         & 270\tpm3.5\tpm35             &   269.6 \\
$K^+$           & 46.7\tpm1.5\tpm7.0           &   46.2 \\ 
$K^-$           & 40.5\tpm2.3\tpm6.1           &   42.9 \\
$p$             & 28.7\tpm0.9\tpm4.0           &   29.6 \\
$\bar{p}$       & 20.1\tpm1.0\tpm2.8           &   20.6 \\
$\Lambda$       & 17.3\tpm1.8\tpm2.8           &   15.9 \\
$\bar{\Lambda}$ & 12.7\tpm1.8\tpm2.0           &   11.8 \\
\hline
\hline
$V'T^3e^{-0.7/T}$ & \multicolumn{2}{|c|}{8.07\tpm0.11} \\
$T$ [MeV]        & \multicolumn{2}{|c|}{158.0\tpm5.9} \\
$\mu_B$ [MeV]    & \multicolumn{2}{|c|}{33.5\tpm17.8} \\
$\gamma_S$       & \multicolumn{2}{|c|}{1.24\tpm0.22} \\
$\chi^2/dof$     &  \multicolumn{2}{|c|}{0.5/4}        \\
\EperNt [GeV]    &  \multicolumn{2}{|c|}{0.97\tpm0.06} \\
\hline
\end{tabular}
\caption{{\bf Top panel:} Rapidity densities around $|y|=0$ in 
central Au-Au collisions at 
\sqrtsnnt=130 GeV measured by the PHENIX collaboration. 
All data taken from~\cite{Adcox:2003nr} except
lambdas from~\cite{Adcox:2002au}. No weak decay corrections have been 
applied to any 
of the particle species. Errors quoted are statistical + systematic
and are taken into account in quadrature in the analysis. 
{\bf Bottom panel:} The statistical hadronization model best fit parameters
at chemical freeze-out and the mean energy per particle
determined from the above experimental data.
\label{phenix130}}
\end{center}
\end{table}

We have repeated the analysis with \tgs~fixed to unity both with the 
STAR and PHENIX rapidity densities at \sqrtsnnt=130 GeV. 
It seems that the minimum of the $\chi^2$ distribution is relatively flat 
in the case of the PHENIX data and the rapidity densities can be described
well ($\chi^2/dof=2.0/3$) with the statistical hadronization model in 
absolute chemical equilibrium as well. The other resulting chemical 
freeze-out parameters are virtually the same as in Table \ref{phenix130}.
Setting \tgs=1 with the STAR rapidity densities leads to somewhat
larger temperature $T=169.1\pm4.4$ MeV while baryon chemical potential
is essentially unaffected. The anomalous behavior of strangeness
phase-space occupancy factor at \sqrtsnnt=130 GeV is not yet well 
understood and the issue deserves further consideration.

\section{Discussion and Conclusions}

The resulting chemical freeze-out parameters can be compared with the 
ones we have determined in central heavy-ion collisions
at lower beam energies at AGS and 
SPS~\cite{Becattini:2005xt}. One should notice that the previous analyses
have been performed with full phase-space multiplicities while the 
statistical hadronization model analysis at RHIC employs rapidity densities 
measured in a limited rapidity window around mid-rapidity. Nevertheless, 
statistical hadronization model parameters in central Au-Au collisions
at RHIC seem to be in good agreement with our interpolating curves for 
statistical hadronization model parameters determined at lower beam 
energies, see Figure 1. The solid line indicates the beam energy region 
in which the interpolating curves are determined while the dashed part is
an extrapolation to RHIC energies. For the explicit formulas and details, 
see~\cite{Becattini:2005xt}.

Figure 2 shows the strangeness phase-space occupancy factor~\tgs~as a 
function of the center-of-mass energy of a colliding nucleon pair in 
central heavy-ion collisions. One can see that strangeness seems to reach
absolute chemical equilibrium in central heavy-ion collisions at RHIC 
energies only. From the right panel of Figure 2 one can see that the RHIC
systems at 200 and 130\agevt seem to follow the chemical freeze-out 
curve~\EperNt~= 1 GeV~\cite{Cleymans:1998fq}.

%==========================================================================
\begin{figure}[!ht]
\begin{tabular}{cc}
\includegraphics[scale=0.7]{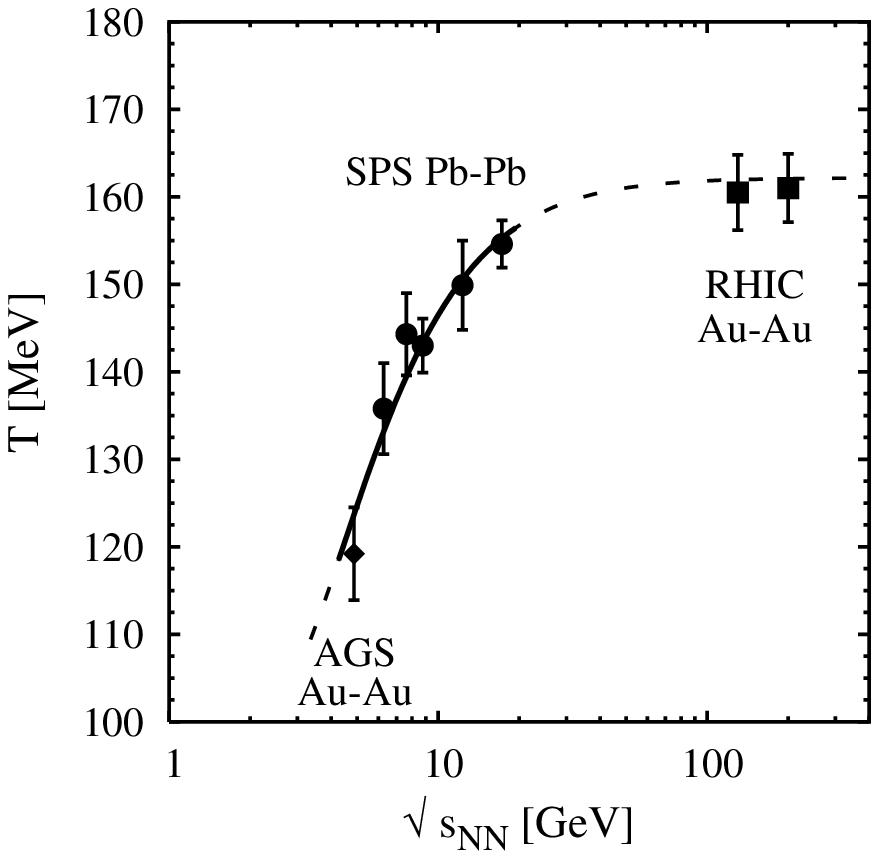} & 
\hspace{0.15cm} 
\includegraphics[scale=0.7]{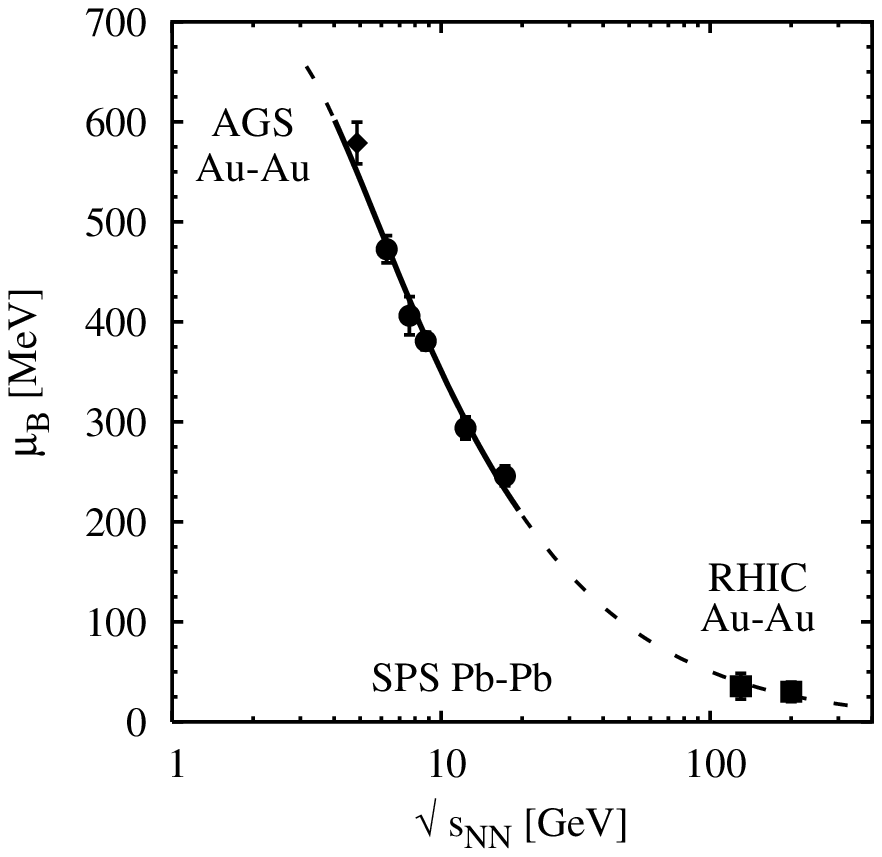} \\
\end{tabular}
\caption{Chemical freeze-out temperature ({\bf LEFT}) and baryon chemical 
potential ({\bf RIGHT}) as a 
function of the center-of-mass energy of a colliding nucleon pair in various 
central collision systems. The lines are our interpolating curves for 
 central heavy-ion collisions~\cite{Becattini:2005xt}.}
\label{temperature}
\end{figure}
%============================================================================
%
%==========================================================================
\begin{figure}[!ht]
\begin{tabular}{cc}
\includegraphics[scale=0.7]{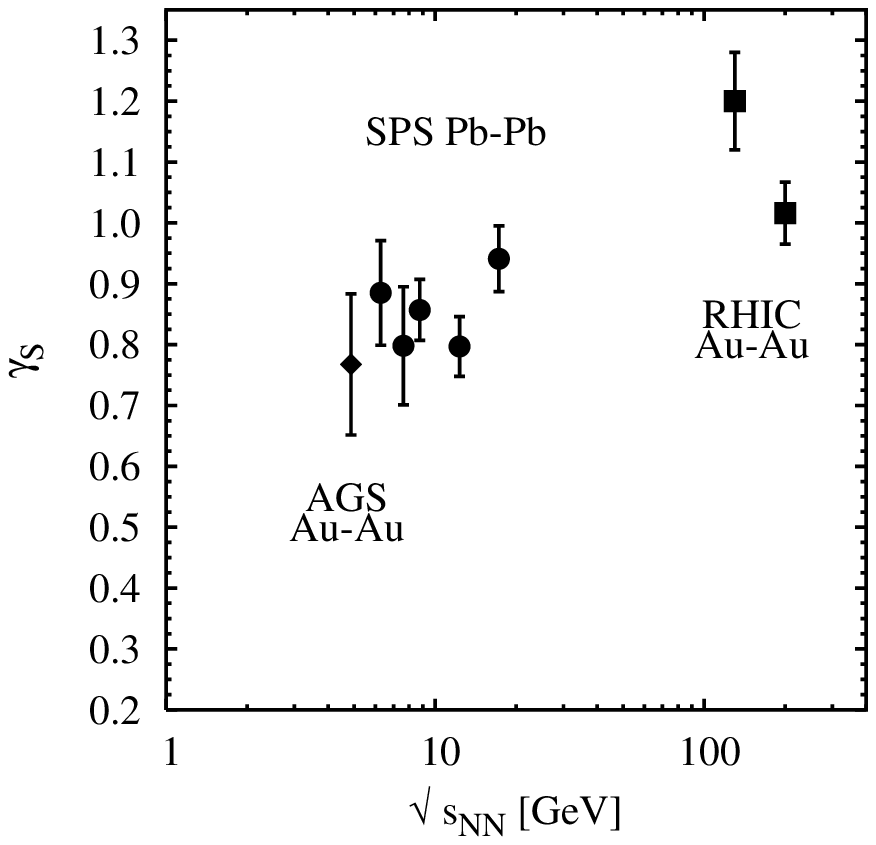} & 
\hspace{0.15cm}
\includegraphics[scale=0.7]{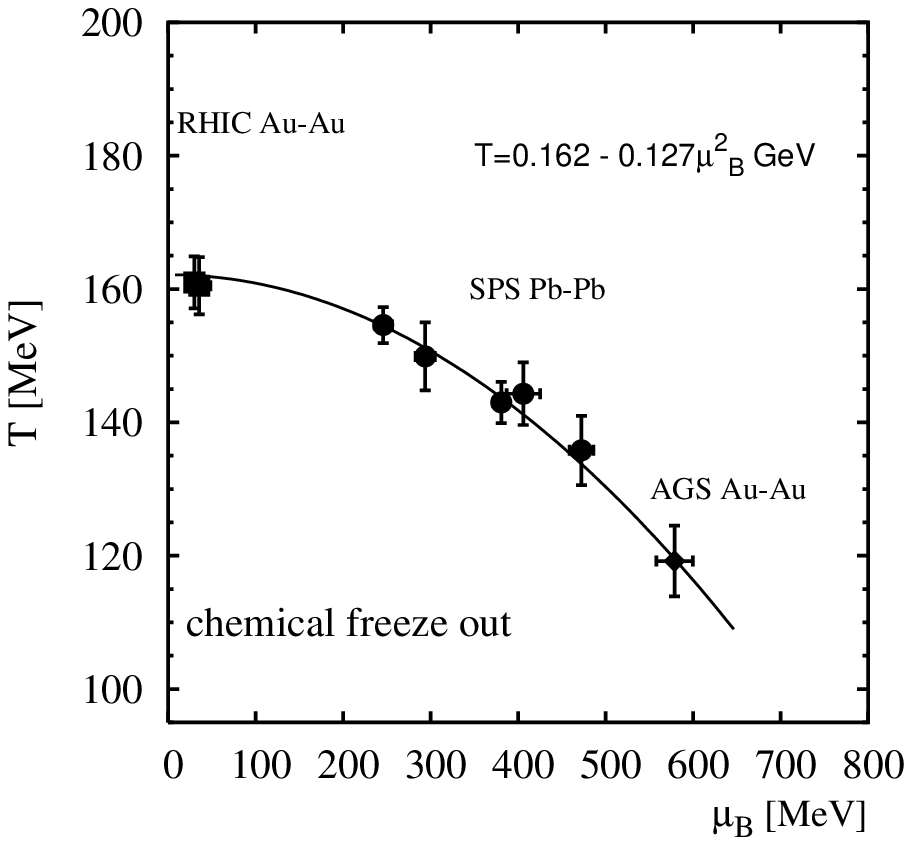} \\
\end{tabular}
\caption{{\bf LEFT:} Strangeness under-saturation parameter $\gamma_S$ as a 
function of the center-of-mass energy of a colliding nucleon pair in various 
central collision systems.\newline
{\bf RIGHT:} The chemical freeze-out curve in central heavy-ion collisions.
The line represents the condition \EperNt\ = 1 GeV.}
\label{chemFO}
\end{figure}
%============================================================================
%

%%%%%%%%%%%%%%%%%%%%%%%%%%%%%%%%%%%%%%%%%%%%

%%%%%%%%%%%%%%%%%%%%%%%%%%%%%%%%%%%%%%%%%%%
\end{document}